# Ultrafast chiral precession of spin and orbital angular momentum induced by circularly polarized laser pulse

Junjie He[1], Shuo Li[2], Thomas Fruenheim[3]

[1]*Faculty of Science, Department of Physical and Macromolecular Chemistry, Charles University, Prague 12843, Czech Republic*

[2]*Institute for Advanced Study, Chengdu University, Chengdu 610100, P. R. China*

[3]*Bremen Center for Computational Materials Science, University of Bremen, Bremen 28359, Germany*

*Email:* junjie.he@natur.cuni.cz, shuoli.phd@gmail.com

**ABSTRACT**

Despite spin (SAM) and orbital (OAM) angular momentum dynamics are well-studied in demagnetization processes, their components receive less focus. Here, we utilize the non-collinear spin version of real-time time-dependent density functional theory (rt-TDDFT) to unveil significant *x* and *y* components of SAM and OAM induced by circularly left ($\sigma^+$) and right ($\sigma^-$) polarized laser pulse in Ferromagnetic Fe, Co and Ni. Our results show that the magnitude of OAM is an order of magnitude larger than that of SAM, highlighting a stronger optical response from the orbital degrees of freedom of electrons compared to spin. Additionally, we observe a marked dependency of the oscillations of the *x* and *y* components on the helicity of the pulse. Intriguingly, $\sigma^+$ and $\sigma^-$ pulses induce chirality in the precession of SAM and OAM, respectively, with clear associations with laser frequency and duration. Finally, we demonstrate that the time scale of OAM and SAM precession occurs even earlier than that of the demagnetization process and OISTR effect. Our investigation provides detailed insight into the dynamics of SAM and OAM of electron during and shortly after a circularly polarized laser excitation.

**Introduction**

Beaurepaire *et al.*[1] made a groundbreaking discovery, showcasing ultrafast demagnetization of ferromagnetic nickel within a subpicosecond timescale through femtosecond laser pulses— a process three orders of magnitude faster than that achievable with magnetic fields alone. This discovery not only has profound implications for fundamental science but also heralds potential breakthroughs in technological applications.[2, 3] With advancements in generating ultrashort laser pulses, the timescale for spin manipulation has now reached femtosecond and even attosecond domains.[4–7] Recently, Dewhurst *et al.*[8] proposed a new mechanism for ultra- fast spin manipulation, termed the optically induced intersite spin transfer (OISTR) effect, which has since received experimental corroboration.[9–13] The OISTR effect demonstrates that optical excitation can coherently and efficiently redistribute spins among distinct magnetic sublattices within tens to a mere few femtoseconds, positioning it as the most rapid method for controlling spin in magnetic materials.[7]

While the OISTR effect originates from light- induced spin-dependent charge excitation on femtosecond timescales and predominantly emphasizes the magnitude of spin transfer and its consequent demagnetization,[7, 8] it does not extensively address the changes in spin component. This leads to pertinent questions: (i) How does the laser pulse influence the spin angular momentum (SAM) components during the OISTR effect? (ii) What role does the orbital angular momentum (OAM) play in the OISTR process? Addressing these questions is of paramount importance to gain a comprehensive understanding of the intricacies involved in the OISTR effect and its potential implications for ultrafast magnetization phenomena.

Spin-orbit (SO) interactions play a fundamental role in the dynamics of demagnetization processes. This relativistic interaction breaks the conservation of the total electronic SAM (defined as **S**), potentially leading to a transfer between SAM and OAM (defined as **L**).[14, 15] Such transfers are critical processes that determine the speed of ultrafast magnetization phenomena. However, despite its paramount importance, the experimental evidence on the interchange between SAM and OAM re- mains ambiguous, as highlighted by studies using magnetic circular dichroism (MCD) and x-ray absorption spectra (XAS).[16–20] For instance,

Boeglin *et al.*[21] indicate that OAM might change even faster than spin in experiments. Conversely, recent theoretical studies offer insights supporting the transfer of SAM into OAM.[22]

While much of the research has primarily concentrated on quantifying the transfer of SAM and OAM—demonstrating an initial angular momentum flow of **L** and **S**, which holds potential for experimental observation.[19–21] However, the temporal evolution of the components of SAM and OAM, especially the *x* and *y* components, have been overlooked for a long time. The components of SAM could be brought about by photo- induced spin-polarized currents and the OAM of electrons is gaining prominence in the realm of *orbitronic* devices.[23–25] Pivoting back to the core issue of demagnetization dynamics, several important questions remain unanswered: (i) How do the directions of SAM and OAM evolve during the demagnetization process? (ii) What is the time scale for components of SAM and OAM (iii) How do their respective components of angular momentum interact and transfer via spin-orbit coupling (SOC)? Addressing these questions is crucial for a comprehensive understanding of the microscopic processes underlying demagnetization dynamics.

To address the aforementioned issues, in this work, we present a detailed study of the time-dependent components of SAM and OAM dynamics under the influence of a circularly polarized laser pulse, utilizing first-principles calculations. We proposed the concept of chirality of spin and orbital precession and subsequently dissect the disparity in precession angles between OAM and SAM. Furthermore, we explored the influence of laser parameters on the precession for OAM and SAM.

We employ the real-time time-dependent density func- tional theory (rt-TDDFT) to study the spin and orbital dynamics of Fe, Co and Ni metals. Previous applications of rt-TDDFT have successfully elucidated ultrafast spin dynamics in metals, Heusler compounds, metallic alloying, and 2D magnets.[7–10, 26–29] The foundation of rt- TDDFT is the Runge–Gross theorem,[30] which posits that a time-dependent external potential uniquely de- fines the functional of time-dependent density, contin- gent upon the initial state. In a fully non-collinear spin- dependent version, time-dependent Kohn–Sham (KS) orbitals are Pauli spinors governed by the Schrodinger equation (refer to details in Ref.[31]):

$$i\frac{\partial \psi_j(\mathbf{r}, t)}{\partial t} = \left[\frac{1}{2}\left(-i\nabla + \frac{1}{c}\mathbf{A}_{\text{ext}}(t)\right)^2 + v_s(\mathbf{r}, t)\right.$$

$$\left. + \frac{1}{2c}\sigma \cdot \mathbf{B}_s(\mathbf{r}, t) + \frac{1}{4c^2}\sigma \cdot (\nabla v_s(\mathbf{r}, t) \times -i\nabla)\right]\psi_j(\mathbf{r}, t)$$

where $\mathbf{A}_{\text{ext}}(t)$ and $\sigma$ represent a vector potential and Pauli matrices. The KS effective potential $v_s(\mathbf{r},t) = v_{ext}(\mathbf{r},t) + v_H(\mathbf{r},t) + v_{xc}(\mathbf{r},t)$ can be decomposed into the external potential $v_{ext}$, the classical Hartree potential $v_H$, and the exchange-correlation (XC) potential $v_{xc}$, respectively. The KS magnetic field can be written as $\mathbf{B}_s(\mathbf{r},t) = \mathbf{B}_{\text{ext}}(\mathbf{r},t) + \mathbf{B}_{\text{xc}}(\mathbf{r},t)$, where $\mathbf{B}_{\text{ext}}$ and $\mathbf{B}_{\text{xc}}$ represent the magnetic field of the applied laser pulse plus possibly an additional magnetic field and XC magnetic field, respectively. The last term in Eq. (1) stand for the SOC. We only time propagate the electronic system while keeping the nuclei fixed.

All calculations were employed by a fully non-collinear spin version[31] of rt-TDDFT by implemented through the full-potential augmented plane-wave ELK code.[32] We utilized a 12 × 12 × 1 regular mesh in k-space, and the rt-TDDFT simulations were conducted with a time step of $\Delta t$= 0.1 a.u. The smearing width was set to 0.027 eV. The laser pulses employed in our analyses were circularly polarized. Additionally, all calculations adhered to the adiabatic local spin density approximation (ALSDA), consistent with methodologies established in previous works.[8–13]

The circularly polarized fields can be written as $\mathbf{A} = (A_x, A_y, 0)$,

$$A_x(t) = \begin{cases} A_0 \cos(\omega t)\sin\left(\frac{\pi t}{T}\right), & 0 \leq t \leq T \\ 0, & \text{otherwise} \end{cases} \quad (2)$$

$$A_y(t) = \begin{cases} A_0 \cos(\omega t - \varphi)\sin\left(\frac{\pi t}{T}\right), & 0 \leq t \leq T \\ 0, & \text{otherwise} \end{cases} \quad (3)$$

where $\varphi$ is the polarization angle and $\varphi = \pm 90°$ are used for circularly σ+ and σ− polarized laser pulses. The second sinusoidal term of Eq. (2) and Eq. (3) constitutes the temporal envelope of the pulse, in which T is a duration.

We first consider a multilayer system of ferromagnetic Fe, Co, and Ni, consisting of a total of six monolayers (ML). Subsequently, a circularly polarized $\sigma^+$ and $\sigma^-$ pulse with a duration (fwhm) of 9.68 fs, 1.63 eV (395THz) frequency and fluence of 7.1 mJ/cm$^2$ was employed to excite the multilayer system. While the time scale for this laser pulse aligns with that employed in the OISTR experiment as referenced in Ref.[7], it notably remains considerably shorter than the durations typically utilized in experiments probing the distinct responses of SAM and OAM dynamics. The ground-state SAM and OAM of the top layer of Co are presented in FIG. 1, accompanied by the temporal evolution of these moments under the influence of $\sigma^+$ and $\sigma^-$ pulse. We exclusively showcase the SAM and OAM dynamics in the top ML due to the similar properties exhibited by the other ML. The other ML exhibits similar properties.

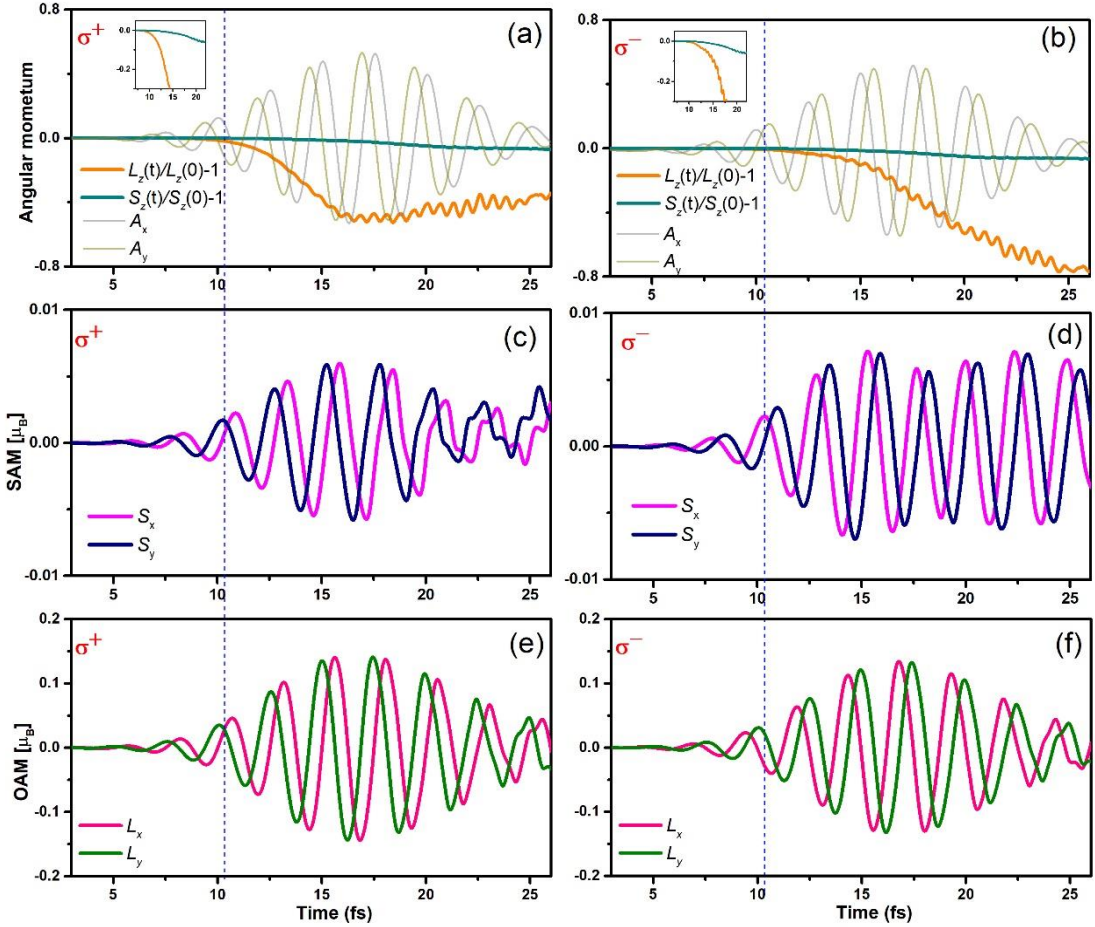

FIG. 1. SAM and OAM dynamics are presented for the *x*, *y*, and *z* components under $\sigma^+$ and $\sigma^-$ pulse excitations. The pulses are polarized with a full width at half maximum (fwhm) of 9.68 fs, central

frequency of 1.63 eV, and an incident fluence of 3.55 mJ/cm$^2$. (a) Normalized orbital angular momentum, $L_z(t)/L_z(0)-1$, and spin angular momentum, $S_z(t)/S_z(0)-1$, are depicted for Co. The light gray and yellow oscillating lines correspond to the $A_x$ and $A_y$ components of the vector potential of the pump pulse, respectively. Time-dependent $x$ (magenta line) and $y$ (blue line) components of SAM dynamics under the $\sigma^+$ (c) and $\sigma^-$ (d) pulses. For SAM, $x$ (pink line) and $y$ (olive line) components under the $\sigma^+$ (e) and $\sigma^-$ (f) pulse excitations are shown.

We now turn our attention to the $x$ and $y$ components (defined as $L_x$, $L_y$, $S_x$ and $S_y$) of both SAM and OAM. In their ground states, $L_x$, $L_y$, $S_x$ and $S_y$ are minimal values. OAM generally quenched due to the motion of electrons in lattice. Interestingly, our results unequivocally demonstrate that a circularly polarized laser pulse induces substantial enhancements in the $x$ and $y$ components of both SAM and OAM for Fe, Co, and Ni, as depicted in FIG. 1c-f. (refer to FIG. S1 and S2 in the supporting materials for Fe and Ni) These metals man- ifest analogous responses when subjected to $\sigma^+$ and $\sigma^-$ pulses. From FIG. 1, We generally observed $L_x$, $L_y$, $S_x$ and $S_y$ demonstrate significant rapid oscillations associated with the frequency of laser ($f_{laser}$). And $L_x$, $L_y$, $S_x$ and $S_y$ show a general increase in the amplitude of oscillation after laser pulse excitation peaking at the culmi- nation of the pump pulse. Subsequently, the oscillations begin to decay and ultimately leading to disorganized behavior following the pulse concludes. For OAM, the decay for amplitudes of $L_x$ and $L_y$ oscillation induced by both $\sigma^+$ and $\sigma^-$ pulses are nearly identical. However, for SAM, the $S_x$ and $S_y$ manifest strong helicity-dependent oscillation behaviour. Specifically, $\sigma^+$ induced oscillated amplitude of $S_x$ and $S_y$ quickly decay, while those induced by $\sigma^-$ exhibit a notably slower decay. Importantly, we note that the OAM amplitudes are approximately 20 times larger than those of SAM, demonstrating that the orbital degree of freedom of electrons have stronger optical response than that of spin. FIG. 1 clearly illustrates that the oscillations of the $x$ and $y$ components precede the reduction of $z$-component of both SAM and OAM. These results indicate that the $x$ and $y$ components play an important role in the early stages of demagnetization dynamics.

The *x* and *y* components of both spin and orbital an- gular momenta for Fe, Co and Ni exhibit a regular oscillation, strongly suggesting an optical helicity-driven precession. FIG. 2 provides visual evidence of this phenomenon, (refer to FIG. S3 and S4 in the supporting materials for Fe and Ni) where $\sigma^-$ light induces a regular left-handed (LH) helix, while $\sigma^-$ light induces a corresponding right-handed (RH) helix in the *x* and *y* components of spin and orbital angular momentum. Notably, these helices originate from the origin and steadily in- crease in amplitude until the laser pulse reaches its peak intensity. Subsequently, the helices tend to return to the origin: we also can observe the oscillatory decay in the *x* and *y* components of spin and orbital angular momenta as shown in FIG. 1. Furthermore, our observations indicate that $\sigma^+$ pulse induces a larger amplitude of precession in spin compared to the precession induced by $\sigma^-$ pulse. That means the spin dynamics show the optical helicity-dependent precession. We can also see that the amplitude of orbital for precession is significantly larger than that of spin, which suggest that OAM will has the larger obliquity of precession than SAM. These findings reveal physical pictures of early dynamics of SAM and OAM under the influence of circularly polarized light, which is important for understanding ultrafast demagnetization processes in magnetic materials.

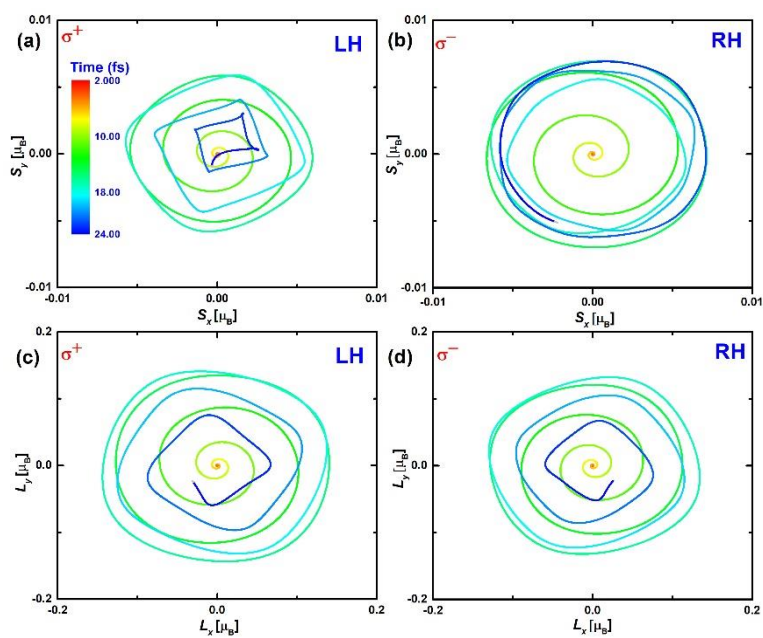

FIG. 2. Left-handed (LH) and right-handed (RH) precession of SAM and OAM induced by circularly polarized pulse. Pan- els (a) and (b) depict the LH and RH precession of SAM under $\sigma^+$ and $\sigma^-$ pulses, respectively. Panels (c) and (d) illustrate the LH and RH precession of OAM under $\sigma^+$ and $\sigma^-$ pulses, respectively. The color maps indicate the time scale from the start to the end of the pulse.

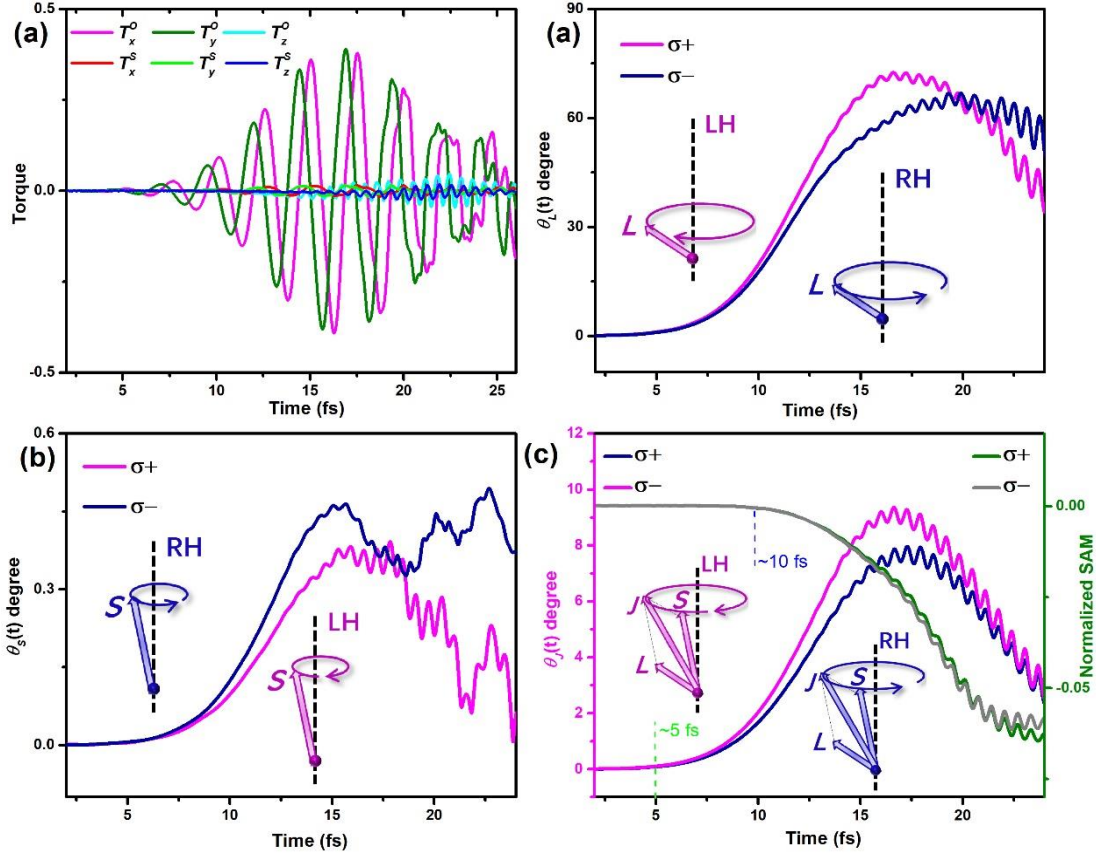

FIG. 3. Laser-induced torques and obliquity of precession for **L**, **S** and **J**. (a) direction-resolved torque for *x*, *y*, and *z* components of SAM and OAM. Time-dependent dynamics of precession obliquity for (b) SAM $\theta_S$, (c) OAM $\theta_L$ and (d) total angular momentum $\theta_J$ induced by $\sigma^+$ (purple line) and $\sigma^-$ (blue line) laser pulse. Inserted schematic illustrations repre- sent the $\sigma^+$ and $\sigma^-$ laser pulse induced left-handed (LH) and right-handed (RH) precession of **L**, **S** and **J**, respectively. The green and blue dash lines with marks at 5 fs and 10 fs denote the estimated initial times for precession and demagnetization, respectively.

The precession of both **L** and **S** results from a torque (**T**) induced by the laser, which can be written as $\mathbf{T}_S=dS/dt$ and $\mathbf{T}_O=dL/dt$ for spin and orbital, respectively. The direction-dependent torque is plotted as a function of time in FIG. 3a. The pulse induces significant large torque for in *x* and *y* components of SAM, while *z*-component of OAM and all directions of SAM exhibit minimal torque. This observation aligns with the aforementioned results for the pronounced magnitudes of $L_x$ and $L_y$ in FIG. 1. The *x* and *y* components of torque for both **L** and **S** regularly oscillate, corresponding to the precession behaviour of **L** and **S**. Subsequently, we further calculated time-dependent obliquity dynamics as illustrated in FIG. 3. For **L**, during early time from 0 fs to 8 fs, the obliquity ($\theta_L$) of OAM presession exhibits slowly increase, then from 8 fs to 18 fs exponentially enhance to maximum saturate value (around 70°). The helicity-dependent dynamics of the $\theta_L$ can also be clearly seen in FiG. 3b: the enhancement induced by $\sigma^+$ surpasses that induced by $\sigma^-$. However, the $\sigma^+$ and $\sigma^-$ only induced a minute spin obliquity ($\theta_S$). FIG. 3c demonstrates that the maximum saturate value of $\theta_S$ is around 0.5°. Despite $\sigma^+$ induces a larger $\theta_S$ angle than $\sigma^-$, the $\theta_S$ angle is two orders of magnitude smaller than $\theta_L$. Consequently, $\theta_S$ is unlikely to substantially influence the obliquity ($\theta_J$) of total angular momentum as evident in FIG. 3d. Notably, the trends for $\theta_J$ and $\theta_L$ exhibit remarkably similarity, with both $\sigma^+$ induced $\theta_J$ and $\theta_L$ are larger than that of $\sigma$-. On the other hand, even though the $\theta_L$ is considerable, but the magnitude of **L** is markedly smaller than counterpart of **S**, leading to a relatively moderate $\theta_J$ value according to vector sum. Overall, the difference between the $\theta_L$ and $\theta_J$ from two helicities is much more pronounced for $\sigma^+$ pulses as compared to $\sigma^-$ pulses. We found that the initiation time for angular momentum precession (beginning at around 5 fs) is evidently earlier than that of demagnetization (comencing around 10 fs).

Next, we will examine how changing the laser param- eters affects the dynamics of SAM and OAM. Firstly, if we look at a longer pulse with fwhm = 26.6 fs, the SAM and the OAM will have precession in a longer time as shown in Figure 4. In FIG. 1, we can clearly see that the frequencies ($\omega_p$) of SAM and OAM precession essentially coincide with the applied laser frequency ($f_{laser}$) of 395 THz (1.63 eV). Therefore, we will investigate whether the $\omega_p$ of SAM and OAM depend on the $f_{laser}$. We consider different incident $f_{laser}$, including 66, 197, 395, 658, 1316 THz, where the fluence and duration of the laser were fixed as shown in FIG. 4. Here, the determi- nation of $\omega_p$ for SAM and OAM is accomplished through Fourier transformation of their respective temporal dy- namics. For FIG. 4a, as $f_{laser}$ increases, we can observe that both $\omega_p$ of **L** and **S** increase linearly, and the $\omega_p$ for **S** and **L** are almost identical. Employing a linear regression analysis, it is established that the precession frequencies for **S** and **L** are in direct proportion to the incident $f_{laser}$.

To further analyse the frequency dependence of SAM and OAM dynamics, we calculated the obliquity ($\theta_S$ and $\theta_L$) of SAM and OAM within different $f_{laser}$ ranging from 66 Thz, to 1316 THz (See FIG. 4dc). Our finding show that the higher $f_{laser}$ laser pulse generally leads to a notable reduction in $\theta_S$ and $\theta_L$, while conversely, lower $f_{laser}$ laser pulses tend to yield larger values of $\theta_S$ and $\theta_L$. Notably, for OAM, as the $f_{laser}$ reaches 1316 THz, the $\theta_L$ gradually increases from 0° to about 10°, and at this frequency, it can give rise to a $\theta_L$ approaching 90°. Since the $f_{laser}$ is directly proportional to the $\omega_p$ of SAM and OAM, lower frequency lasers result in a slower $\omega_p$, thereby leading to an increase in the precession angle. Such behavior parallels that of classical gyroscope precession, where a decrease in precession an- gular frequency lead to an increase in precession angle. consequently, when a sufficiently high $f_{laser}$ laser pulse is applied, it drives the precession angle to towards 90°, resulting in an unstable precession. In the case of gyroscope precession, an increase in the precession angle or a decrease in $\omega_p$ also lead to an unstable motion in precession. Regarding SAM, we similarly observe a substantial influence of $f_{laser}$ on the $\theta_S$. However, unlike the OAM, the SAM only exhibits a small precession angle. For example, a high $f_{laser}$ of 1316 THz yield a maximum $\theta_S$ of merely 0.05°. While a low $f_{laser}$ can enhance the $\theta_S$ by approximately 30-fold, reaching 1.5°, it still remains smaller in comparison to OAM. Our finding illustrates the $f_{laser}$ can effectively manipulate the precession of both SAM and OAM.

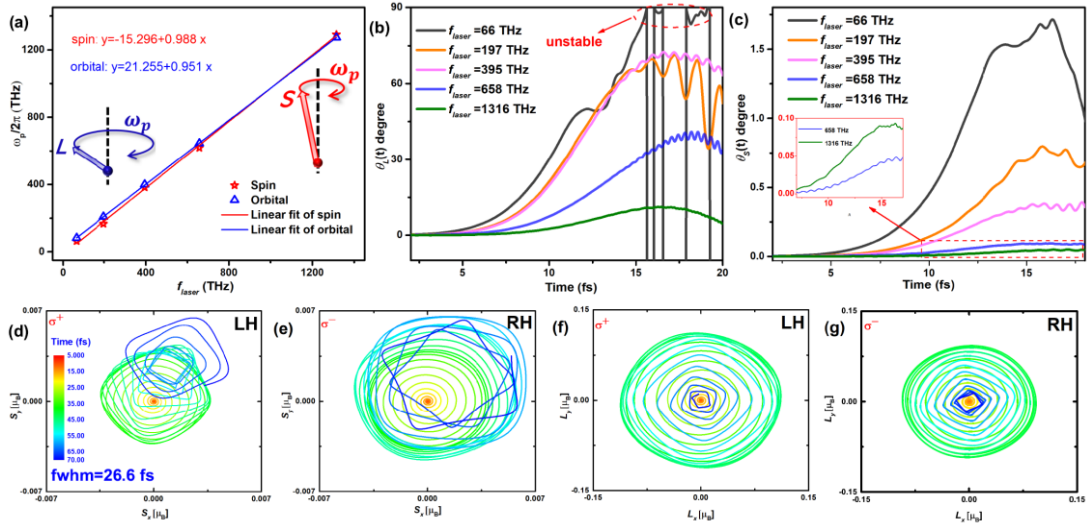

FIG. 4. The dynamics of SAM and OAM affects by changing laser parameters. (a) The relationship between laser frequency ($f_{laser}$) and the frequency ($\omega_p/2\pi$) of precession for SAM and OAM. Inserted schematic illustrations represent the precession angular frequency ($\omega_p$) of **L** and **S**, respectively. Frequency-dependent precession obliquities for (b) $\theta_S$, (c) $\theta_L$ are displayed. The unstable state in low frequency $f_{laser}$=66 THz is also indicated in (b). Panels (d) and (e) depict the LH and RH precession of SAM under $\sigma^+$ and $\sigma^-$ pulses with fwhm=26.6 fs, respectively. Panels (f) and (g) illustrate the LH and RH precession of OAM under $\sigma^+$ and $\sigma^-$ pulses with the same parameter, respectively. The color maps indicate the time scale from the start to the end of the pulse.

Previous studies have primarily focused on analyzing the dynamics of the *z*-component of angular momentum, regarding them as the primary factors contributing to both demagnetization and angular momentum transfer. However, the *x* and *y* components of both SAM and OAM, which have received comparatively less attention, may exert significant influence over magnetization dynamics. Recent theoretical works have showcased distinct spin dynamics in magnetic or non-magnetic materials induced by a circularly polarized laser pulse;[33,34] however, chiral spin precession and oscillated transverse magnetism have not been reported to date. Notably, our findings demonstrate that a circularly polarized laser pulse induces a considerably larger component of OAM compared to SAM, signifying a more robust optical response from OAM. From Figure S5, we can also see that the OAM has a stronger optical response than the SAM as the amplitude of the pulse changes. These results suggest that the angular momentum of a circularly polarized laser may predominantly transfer to the orbital component, driving spin precession through the SOC. The dynamics of $x$ and $y$ components of SAM and OAM in Co are shown for SOC scaled by factors of 1.5 and 2.0. (See Figure S6) It is clear that an increase in SOC leads to an increased SAM but no change in OAM. Moreover, it is worth noting that theoretical calculations have reported the induction of OAM components of electrons through femtosecond laser pulses in Co/Cu(100) interfaces.[35, 36] and Pt films[37] and metallic cluster[38]. Consequently, the significant generation of OAM components by light demonstrates consider- able potential in the emerging field of *orbitronics*.[23, 25]

The OISTR effect proposes that light can directly and coherently interact with spin, representing the fastest means of controlling spin. This is achieved through light- induced spin-selective charge excitation within the sublattice of magnetic materials. Our results demonstrate that the time scale of OAM and SAM precession occurs extremely rapidly (<10 fs), even preceding that of the OISTR effect and demagnetization. This is supported by the corresponding time scale of the $x$ and $y$ components of SAM and OAM, as illustrated in Figure 1c-f and Figure 3. The orientation of SAM and OAM plays a crucial role in comprehending the microscopic mechanisms involved in laser-induced demagnetization. Moreover, it's important to note that our simulations were specifically focused on the early spin dynamics, thus limiting the time scale to within 100 fs. However, in the time scale of 50 to 100 fs, electron-phonon coupling will play a crucial role in influencing the precession of SAM and OAM, leading to a complex angular momentum transfer involving phonons. Recently, Tauchert et al. [39] observed *circularly polarized phonons* or *chiral phonons* in the demagnetization process due to angular momentum transfer from spin systems. The angular momentum transfer between polarized phonon and chirality of SAM and OAM dynamics presents an intriguing open question warranting further investigation.

In summary, we employed rt-TDDFT simulations to explore the SAM and OAM dynamics in ferromagnetic Fe, Co, and Ni, subjected to circularly $\sigma^+$ and $\sigma^-$ polarized laser pulses. We unveiled pronounced *x* and *y* components for both SAM and OAM, with the OAM components exhibiting an order of magnitude larger magnitude than their SAM counterparts. This observation emphasizes a more substantial optical response emanating from the electrons' orbital degrees of freedom in comparison to their spin. Furthermore, we noted a clear dependence of the *x* and *y* of component oscillations on the optical helicity. Intriguingly, $\sigma^+$ and $\sigma^-$ pulses were observed to induce distinct chirality in the precession of SAM and OAM: the $\sigma^+$ pulse results in regular LH helical dynamics, while the $\sigma^-$ pulse fosters a RH helical dynamics of both SAM and OAM. These chiral precession dynamics show a strong correlation with the laser's frequency and duration. Our results reveal an exceptionally rapid precession of OAM and SAM, outstripping the timescale of the demagnetization process and the OISTR effect. Such chirality of spin and orbital dynamics could be important for the circularly polarized phonon in the demagnetization process.

**ACKNOWLEDGMENTS**

J.H. acknowledge the e-INFRA CZ (ID:90140) for providing computational resources and the funding support from MSCA Fellowships CZ–UK with CZ.02.01.01/00/22_010/0002902. S. L. acknowledges the support from National Natural Science Foundation of China (Grant No. 12204069). We also acknowledge ChatGPT for improving readability and language.

**REFERENCES**

[1] E. Beaurepaire, J.-C. Merle, A. Daunois, and J.-Y. Bigot, Phys. Rev. Lett. **76**, 4250 (1996).
[2] A. Kirilyuk, A. V. Kimel, and T. Rasing, Rev. Mod. Phys. **82**, 2731 (2010).
[3] C. D. Stanciu, F. Hansteen, A. V. Kimel, A. Kirilyuk, A. Tsukamoto, A. Itoh, and T. Rasing, Phys. Rev. Lett. **99**, 047601 (2007).
[4] I. Radu, K. Vahaplar, C. Stamm, T. Kachel, N. Pontius, H. Du¨rr, T. Ostler, J. Barker, R. Evans, R. Chantrell, *et al.*, Nature **472**, 205 (2011).
[5] J.-Y. Bigot, M. Vomir, and E. Beaurepaire, Nat. Phys. **5**, 515 (2009).
[6] B. Koopmans, G. Malinowski, F. Dalla Longa, D. Steiauf, M. F¨ahnle, T. Roth, M. Cinchetti, and M. Aeschlimann, Nat. Mater. **9**, 259 (2010).
[7] F. Siegrist, J. A. Gessner, M. Ossiander, C. Denker, Y.-P. Chang, M. C. Schr¨oder, A. Guggenmos, Y. Cui, J. Walowski, U. Martens, *et al.*, Nature **571**, 240 (2019).

[8] J. K. Dewhurst, P. Elliott, S. Shallcross, E. K. Gross, and S. Sharma, Nano Lett. **18**, 1842 (2018).

[9] F. Willems, C. von Korff Schmising, C. Strüber, D. Schick, D. W. Engel, J. Dewhurst, P. Elliott, S. Sharma, and S. Eisebitt, Nat. Commun. **11**, 871 (2020).

[10] M. Hofherr, S. Häuser, J. Dewhurst, P. Tengdin, S. Sak-shath, H. Nembach, S. Weber, J. Shaw, T. J. Silva, H. Kapteyn, *et al.*, Sci. Adv. **6**, eaay8717 (2020).

[11] M. Hofherr, S. Häuser, J. Dewhurst, P. Tengdin, S. Sak-shath, H. Nembach, S. Weber, J. Shaw, T. J. Silva, H. Kapteyn, *et al.*, Sci. Adv. **6**, eaay8717 (2020).

[12] C. von Korff Schmising, S. Jana, K. Yao, M. Hennecke, P. Scheid, S. Sharma, M. Viret, J.-Y. Chauleau, D. Schick, and S. Eisebitt, Phys. Rev. Res. **5**, 013147 (2023).

[13] E. Golias, I. Kumberg, I. Gelen, S. Thakur, J. Gördes, R. Hosseinifar, Q. Guillet, J. Dewhurst, S. Sharma, C. Schüßler-Langeheine, *et al.*, Phys. Rev. Lett. **126**, 107202 (2021).

[14] W. Töws and G. Pastor, Phys. Rev. Lett. **115**, 217204 (2015).

[15] G. Lefkidis, G. Zhang, and W. Hübner, Phys. Rev. Lett. **103**, 217401 (2009).

[16] B. Thole, P. Carra, F. Sette, and G. van der Laan, Phys. Rev. Lett. **68**, 1943 (1992).

[17] P. Carra, B. Thole, M. Altarelli, and X. Wang, Phys. Rev. Lett. **70**, 694 (1993).

[18] C. Chen, Y. Idzerda, H.-J. Lin, N. Smith, G. Meigs, E. Chaban, G. Ho, E. Pellegrin, and F. Sette, Phys. Rev. Lett. **75**, 152 (1995).

[19] N. Bergeard, V. López-Flores, V. Halte, M. Hehn, C. Stamm, N. Pontius, E. Beaurepaire, and C. Boeglin, Nat. Commun. **5**, 3466 (2014).

[20] M. Hennecke, I. Radu, R. Abrudan, T. Kachel, K. Holl-dack, R. Mitzner, A. Tsukamoto, and S. Eisebitt, Phys. Rev. Lett. **122**, 157202 (2019).

[21] C. Boeglin, E. Beaurepaire, V. Halté, V. López-Flores, C. Stamm, N. Pontius, H. Dürr, and J.-Y. Bigot, Nature **465**, 458 (2010).

[22] J. Dewhurst, S. Shallcross, P. Elliott, S. Eisebitt, C. v. K. Schmising, and S. Sharma, Phys. Rev. B **104**, 054438 (2021).

[23] Y.-G. Choi, D. Jo, K.-H. Ko, D. Go, K.-H. Kim, H. G. Park, C. Kim, B.-C. Min, G.-M. Choi, and H.-W. Lee, Nature **619**, 52 (2023).

[24] G. Sala, H. Wang, W. Legrand, and P. Gambardella, Phys. Rev. Lett. **131**, 156703 (2023).

[25] I. Lyalin, S. Alikhah, M. Berritta, P. M. Oppeneer, and R. K. Kawakami, arXiv preprint arXiv:2306.10673 (2023).

[26] S. Sharma, S. Shallcross, P. Elliott, and J. K. Dewhurst, Sci. Adv. **8**, eabq2021 (2022).

[27] J. Dewhurst, S. Shallcross, E. Gross, and S. Sharma, Phys. Rev. Appl. **10**, 044065 (2018).

[28] J. He, S. Li, A. Bandyopadhyay, and T. Frauenheim, Nano Lett. **21**, 3237 (2021).

[29] J. He, S. Li, T. Frauenheim, and Z. Zhou, Nano Lett. **23**, 8348 (2023).

[30] E. Runge and E. K. Gross, Phys. Rev. Lett. **52**, 997 (1984).

[31] K. Krieger, J. Dewhurst, P. Elliott, S. Sharma, and E. Gross, J. Chem. Theory Comput. **11**, 4870 (2015).

[32] J. Dewhurst and S. Sharma, http://elk.sourceforge.net.

[33] Scheid, P., Sharma, S., Malinowski, G., Mangin, S. & Lebegue, S. Nano Lett. 21, 1943–1947 (2021).

[34] Neufeld, O., Tancogne-Dejean, N., De Giovannini, U., Hubener, H. & Rubio, A. npj Comput. Mater. 9, 39 (2023).

[35] O. Busch, F. Ziolkowski, I. Mertig, and J. Henk, Phys. Rev. B **108**, 104408 (2023).

[36] O. Busch, F. Ziolkowski, I. Mertig, and J. Henk, Phys. Rev. B **108**, 184401 (2023).

[37] Hamamera, H., Guimaraes, F. S. M., Dias, M. d. S. & Lounis, S. arXiv preprint arXiv:2312.07888 (2023).

[38] Sinha-Roy, R., Hurst, J., Manfredi, G. & Hervieux, P.-A. ACS photonics 7, 2429–2439 (2020).

[39] S. R. Tauchert, M. Volkov, D. Ehberger, D. Kazenwadel, M. Evers, H. Lange, A. Donges, A. Book, W. Kreuzpaint- ner, U. Nowak, *et al.*, Nature **602**, 73 (2022).

Supplementary Materials "**Ultrafast chiral precession of spin and orbital angular momentum induced by circularly**

Junjie He[1], Shuo Li[2], Thomas Frauenheim[3]

[1] Department of Physical and Macromolecular Chemistry, Faculty of Science,
Charles University, Prague 12843, Czech Republic

[2] Institute of Advanced Study, Chengdu University, Chengdu 610100, China

[3] Bremen Center for Computational Materials Science, University of Bremen, Bremen 28359, Germany

E-mail: junjie.he@natur.cuni.cz; shuoli.phd@gmail.com

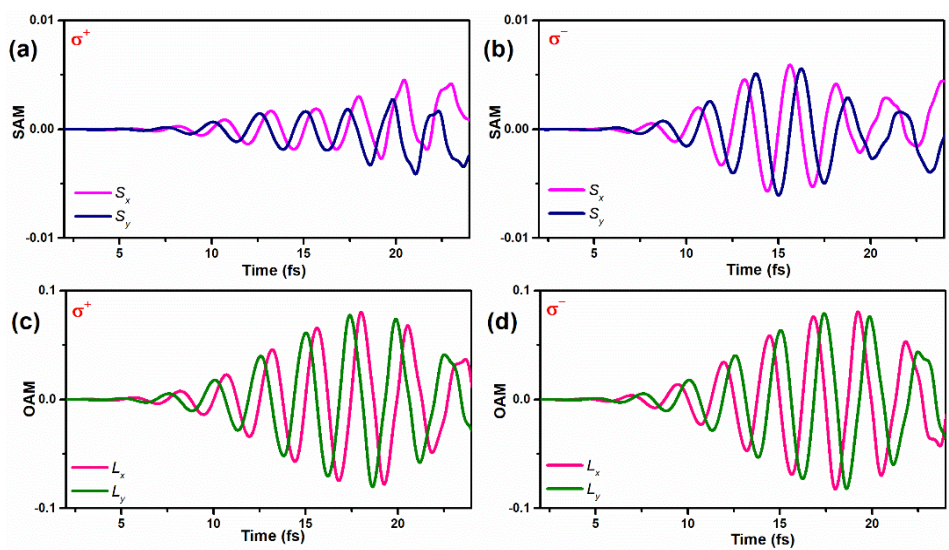

Figure S1: The x and y components of SAM and OAM dynamics for Fe under excitation by $\sigma^+$ and $\sigma^-$ pulse, with the pulses being circularly polarized and characterized by fwhm of 9.68 fs, a central frequency of 1.63 eV, and an incident fluence of 7.1 mJ/cm². Time-dependent x (magenta line) and y (blue line) components of SAM dynamics under the $\sigma^+$ (a) and $\sigma^-$ (b) pulses excitations. The x (pink line) and y (olive line) components of OAM under $\sigma^+$ (c) and $\sigma^-$ (d) pulse excitations.

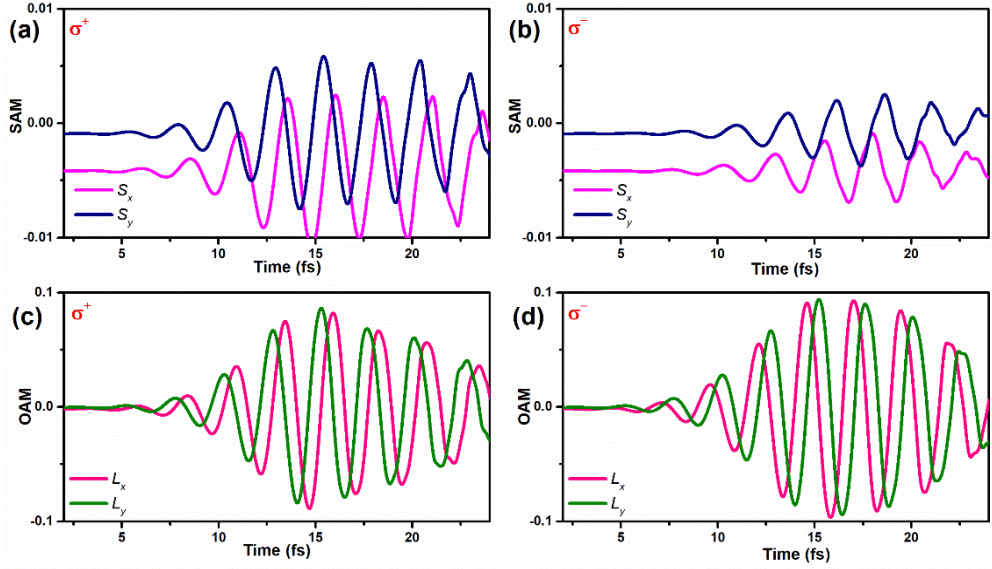

Figure S2: The x and y components of SAM and OAM dynamics for Ni under excitation by $\sigma^+$ and $\sigma^-$ pulse, with the pulses being circularly polarized and characterized by fwhm of 9.68 fs, a central frequency of 1.63 eV, and an incident fluence of 7.1 mJ/cm². Time-dependent x (magenta line) and y (blue line) components of SAM dynamics under the $\sigma^+$ (a) and $\sigma^-$ (b) pulses excitations. The x (pink line) and y (olive line) components of OAM under $\sigma^+$ (c) and $\sigma^-$ (d) pulse excitations.

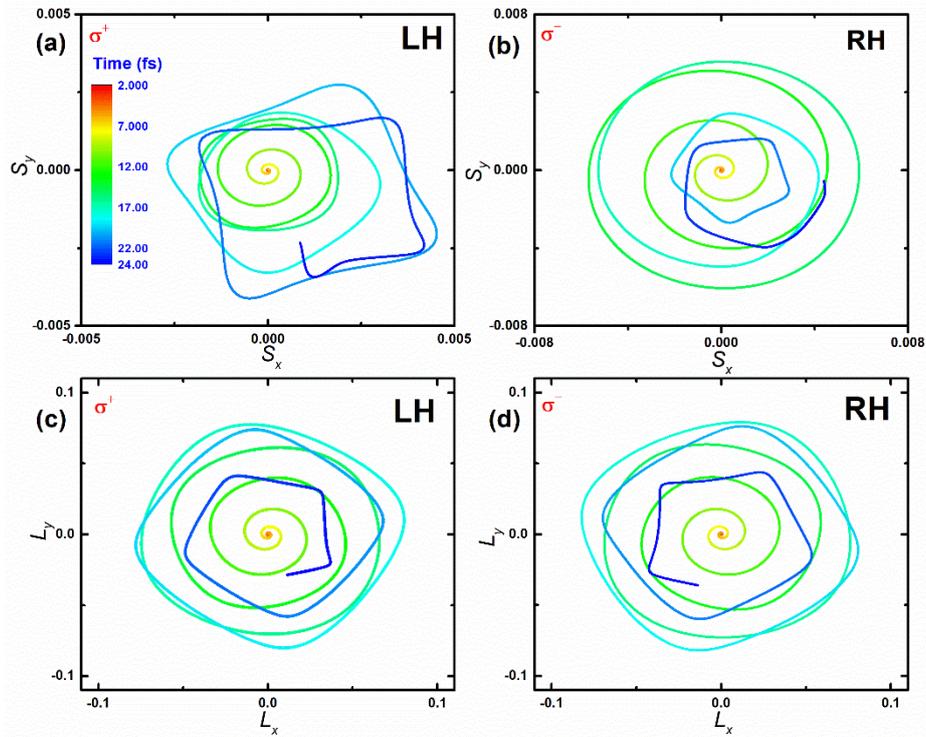

Figure S3: Left-handed (LH) and right-handed (RH) precession of SAM and OAM for Fe

induced by circularly polarized pulse. Panels (a) and (b) depict the LH and RH precession of SAM under $\sigma^+$ and $\sigma^-$ pulses, respectively. Panels (c) and (d) illustrate the LH and RH precession of OAM under $\sigma^+$ and $\sigma^-$ pulses, respectively. The color maps indicate the time scale from the start to the end of the pulse.

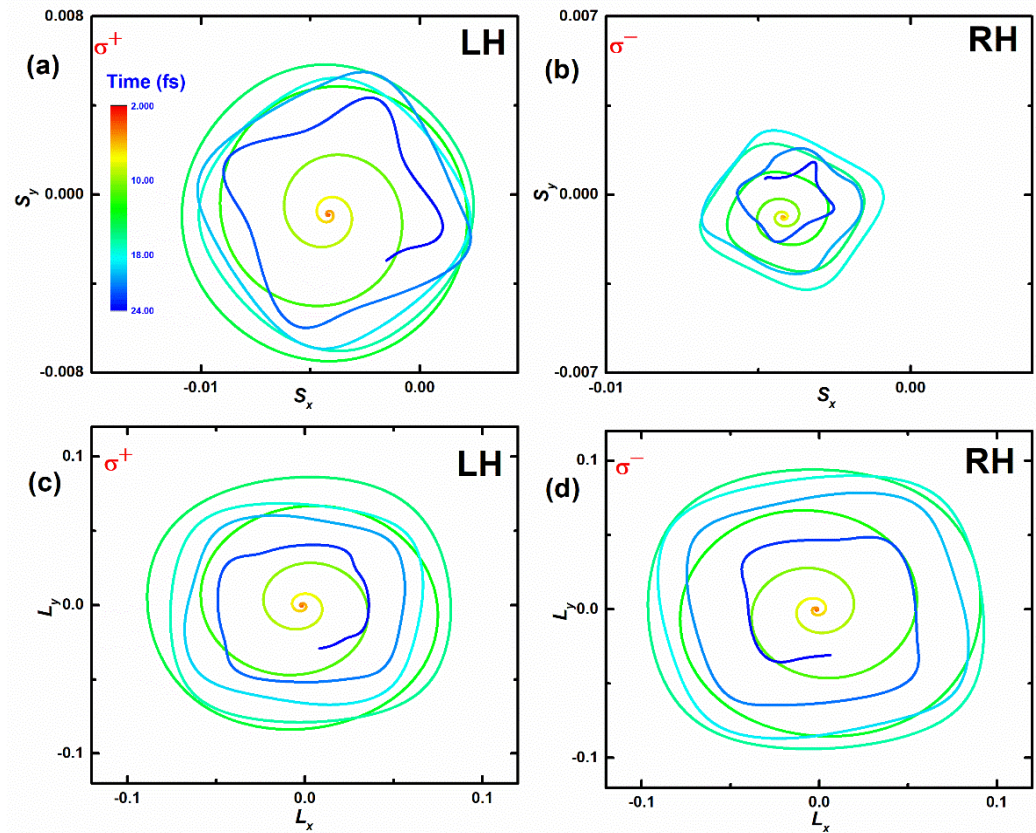

Figure S4: Left-handed (LH) and right-handed (RH) precession of SAM and OAM for Ni induced by circularly polarized pulse. Panels (a) and (b) depict the LH and RH precession of SAM under $\sigma^+$ and $\sigma^-$ pulses, respectively. Panels (c) and (d) illustrate the LH and RH precession of OAM under $\sigma^+$ and $\sigma^-$ pulses, respectively. The color maps indicate the time scale from the start to the end of the pulse.

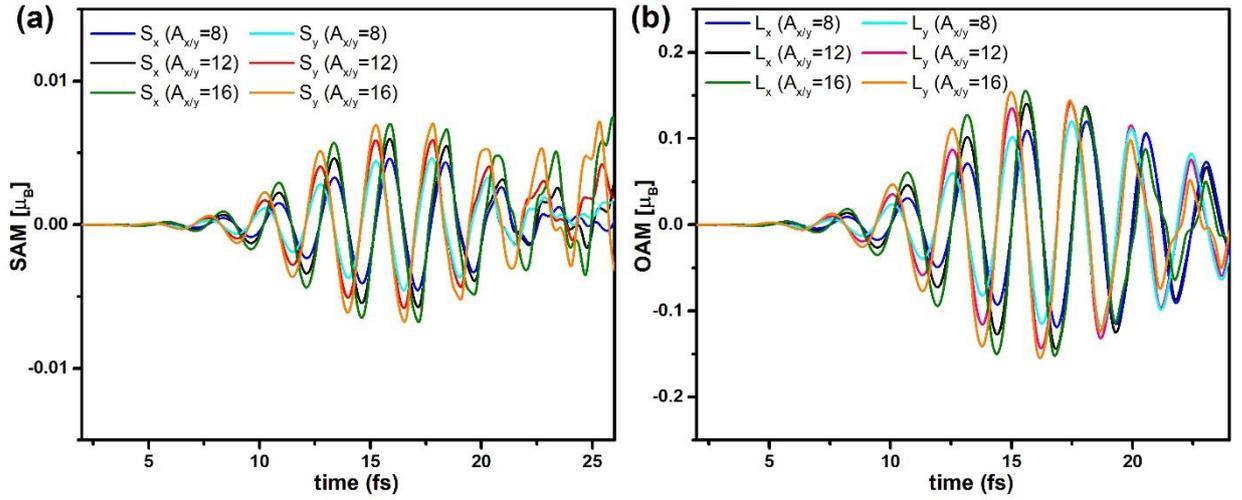

Figure S5. Dependence of the dynamics of *x* and *y* component of SAM (a) and OAM (b) amplitude of pulse. The SAM and OAM dynamics are shown for amplitude with 8, 12 and 16 respectively.

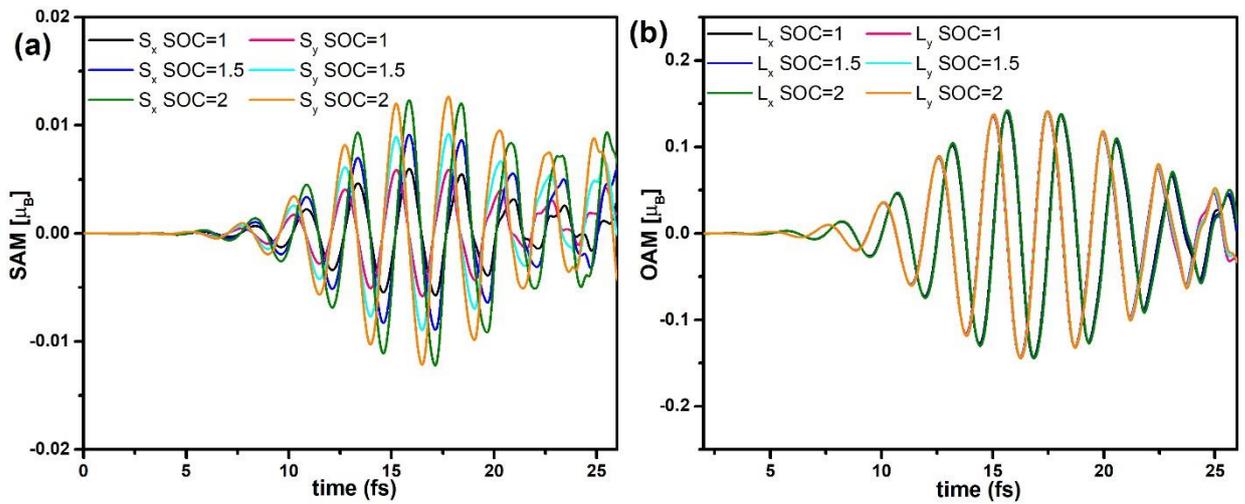

Figure S5. Dependence of the dynamics of *x* and *y* component of SAM (a) and OAM (b) on the spin-orbit coupling (SOC) constant. The SAM and OAM dynamics are shown for SOC scaled by factors of 1, 1.5 and 2.0.

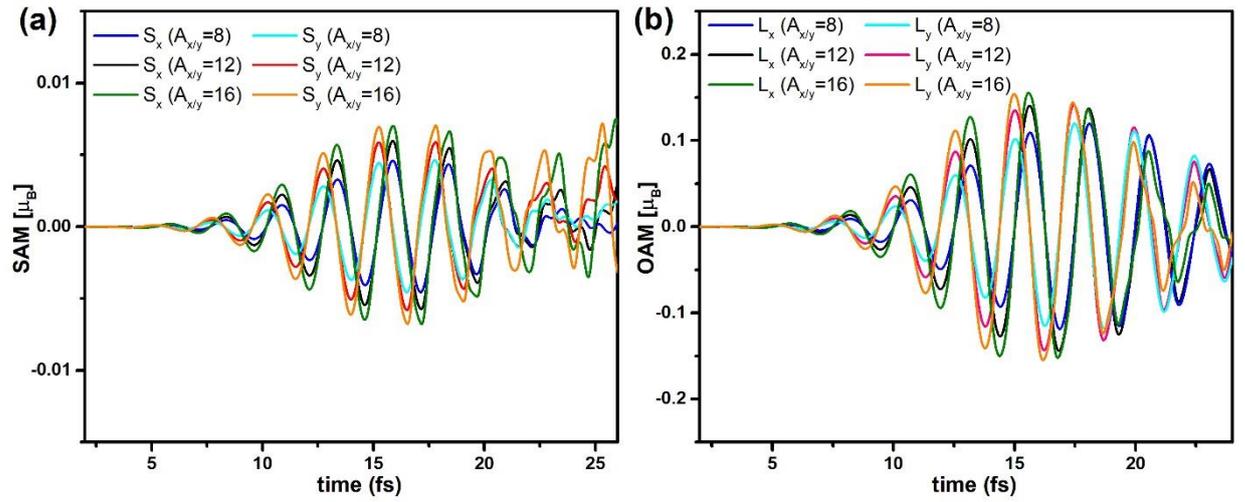

Figure S6. Dependence of the dynamics of *x* and *y* component of SAM (a) and OAM (b) amplitude of pulse. The SAM and OAM dynamics are shown for amplitude with 8, 12 and 16 respectively.